\documentclass[twocolumn,showpacs,amsmath,amssymb, aps, prl, showkeys]{revtex4-1}

\usepackage{graphicx}
\usepackage{natbib}

\begin{document}


\title{Steady-state, cavity-less, multimode superradiance in a cold vapor}


\author{Joel A. Greenberg}
\email[]{jag27@phy.duke.edu}
\affiliation{Department of Physics and the Fitzpatrick Institute for Photonics, Duke University, Durham, NC 27708, USA}
\author{Daniel J. Gauthier}
\affiliation{Department of Physics and the Fitzpatrick Institute for Photonics, Duke University, Durham, NC 27708, USA}


\date{\today}

\begin{abstract}

We demonstrate steady-state, mirrorless superradiance in a cold vapor pumped by weak optical fields. Beyond a critical pump intensity of 1 mW/cm$^2$, the vapor spontaneously transforms into a spatially self-organized state: a density grating forms.  Scattering of the pump beams off this grating generates a pair of new, intense optical fields that act back on the vapor to enhance the atomic organization. We map out experimentally the superradiant phase transition boundary and show that it is well-described by our theoretical model.  The resulting superradiant emission is nearly coherent, persists for several seconds,  displays strong temporal correlations between the various modes, and  has a coherence time of several hundred $\mu$s.  This system therefore has applications in fundamental studies of many-body physics with long-range interactions as well as all-optical and quantum information processing.
  
\end{abstract}


\maketitle

The study of collective light-matter interactions, where the dynamics of an individual scatterer depend on the state of the entire multi-scatterer system, has recently received much attention in the areas of fundamental research and photonic technologies \cite{meiser10b, simon07}. One prominent example of collective behavior is superradiance \cite{sfnote}, where light-induced couplings between initially incoherently-prepared emitters cause the full ensemble to synchronize and radiate coherently \cite{dicke54}.    While  early studies of superradiance focused on collective scattering via the emitters' internal degrees of freedom, recent work demonstrates that formally identical behavior arises through the manipulation of the center-of-mass positions and momenta of cold atoms \cite{slama07, baumann10, nagy10}.

In these studies, an initially uniformly-distributed gas of atoms pumped by external optical fields spontaneously undergoes a transition to a spatially-ordered state under certain circumstances \cite{slama07, baumann10, inouye99, nagy10}.  This  ordering  arises from the momentum imparted to the atoms via optical scattering and can be understood as a form of atomic synchronization: instead of the atoms scattering light individually, the self-assembled density grating  enables the entire ensemble to coherently scatter light as a single entity. The pump beams scatter off this grating and produce new optical fields that act back on the vapor to enhance the grating contrast.   This emergent, dynamical organization can lead to reduced optical instability thresholds \cite{saffman08} and new phenomena \cite{gopal11} that are inaccessible using static, externally-imposed optical lattices \cite{schilke11dfb}.

In order for superradiance to occur, the system must posses sufficient gain and feedback so that synchronization occurs more rapidly than dephasing.  The main dephasing mechanisms are grating washout due to thermal atomic motion and the loss of photons from the interaction volume \cite{slama07}.  One can overcome the effects of thermal motion in free space by working at ultracold temperatures ($T<3$ $\mu$K)  and using optical fields detuned far from the atomic resonance in order to avoid recoil-induced heating \cite{inouye99, yoshikawa05}.  Multi-mode superradiance has been observed in such systems \cite{inouye99}, although the emission is inherently transient because the recoil associated with repeated scattering events eventually destroys the ultracold gas.  Alternatively, placing the atoms in a single-mode cavity effectively increases the coherence time of the system and enables superradiance to occur at temperatures of up to hundreds of $\mu$K \cite{slama07}. Unfortunately, single-mode cavities are incompatible with multi-mode fields, which are necessary for realizing recently-proposed spin glass systems \cite{strack11, gopal11, nagy10} and multi-mode quantum information processing schemes \cite{lassen09, boyer08}.  By using instead a multi-mode cavity and an auxiliary cooling mechanism, recent studies have demonstrated steady-state behavior \cite{robb04e, black03}.  Nevertheless, cavities are more technically challenging to work with than free-space systems and impose additional constraints on the allowed optical frequencies.   

In this Rapid Communication, we demonstrate a collective, superradiant instability that results in the steady-state emission of  multi-mode optical fields in the absence of an optical cavity and without requiring ultracold temperatures. We circumvent dephasing by using a recently-reported light-matter interaction that arises when one works near an atomic resonance. The system's transition to a superradiant state coincides with the onset of global spatial organization in three dimensions through the breaking of continuous translation symmetry in the cold atomic vapor.  We map out the  boundary between the normal and superradiant phases and demonstrate that this transition occurs for intensities as low as $1$ mW/cm$^2$.  Up to $20\%$ of the incident light scatters into the superradiant modes, which display strong temporal correlations and coherence times of several hundred $\mu$s. 

 To realize this instability, we use a magneto-optical trap (MOT) to produce  an anisotropic, pencil-shaped cloud of cold ($T=30$ $\mu$K) $^{87}$Rb atoms \cite{greenberg07}.  The cloud has a length  $L=3$ cm and $1/e$ diameter $W=430$ $\mu$m, which corresponds to a Fresnel number $F=\pi W^2/(\lambda L)\cong6$.  We achieve atomic densities of up to $\eta=3\times10^{10}$ cm$^{-3}$, resulting in resonant optical depths ($OD=\eta \sigma_{23} L$) of $\sim 60$ for the $5S_{1/2}(F=2)\rightarrow5P_{3/2}(F'=3)$ atomic transition, where $\sigma_{23}=7 \lambda^2/20 \pi$ is the effective absorption cross section and $\lambda=780$ nm is the wavelength of light.   
 
\begin{figure}
   \begin{center}  \includegraphics[width=3.2 in]{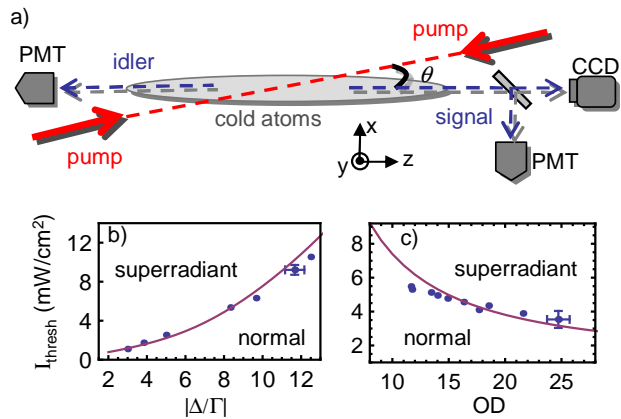}  \end{center} 
   \vspace{-15pt}
  \caption{a) When we shine a pair of weak, counterpropagating pump beams on a pencil-shaped cloud of cold atoms, scattering via the spontaneously-formed density grating produces signal and idler fields propagating along the trap's long axis.  We use a pair of matched photomultiplier tubes (PMTs) and a charge-coupled device (CCD) camera to image the temporal and spatial profile of the generated light, respectively.  Measured (points) and predicted (line) boundary between the uniform (normal) and superradiant  phases as a function of a) detuning (for $OD=20\pm1$) and b) optical depth (for $\Delta/\Gamma=-7$).  The error bars represent typical measurement uncertainties of one standard deviation.}
\label{fig:fig1}
\end{figure}

   After cooling and trapping the atoms in the MOT for 99 ms, we reduce the MOT beams' intensities for up to several seconds (limited by the timing electronics).  During this time, we leave the MOT repump beams and magnetic fields on to ensure that atoms are not pumped into a dark state and to enable continuous cooling and trapping, respectively.  We have verified that the MOT magnetic fields do not affect the superradiance when the MOT beams are off \cite{greenberg11}.   While the MOT beam intensities are reduced, we illuminate the atomic cloud with a pair of pump beams that are intensity-balanced (with single beam intensity $I_p$),  frequency degenerate, and counterpropagate at an angle $\theta=10^\circ$ relative to the trap's long axis (see Fig. \ref{fig:fig1}a).    Unlike in previous studies, we choose detunings $\Delta$ near the $5S_{1/2}(F=2)\rightarrow5P_{3/2}(F'=3)$ atomic transition (\textit{i.e.}, within several natural linewdiths $\Gamma$) and pump beam polarizations  that enable net cooling via the Sisyphus effect \cite{dalibard89}.  
   
    This configuration allows us to exploit a gain mechanism in which  atomic cooling enhances the loading of atoms into the optical lattice formed by the interference of the pump and self-generated optical fields \cite{greenberg11, greenberg12}.   The resulting atomic spatial organization constitutes a density grating, which is phase-matched for coupling the pump and superradiant fields.  Atomic cooling in this compliant optical lattice results in typical atomic temperatures of a few $\mu$K, and the mutual amplification of the optical and density waves via this collective scattering mechanism results in the superradiant instability \cite{inouye99, nagy10, wmnote}.

Because the instability requires that the single-pass gain exceed the intrinsic system loss, superradiance only occurs when $I_p$ exceeds a threshold pump intensity $I_{thresh}$.  In general, the value of $I_{thresh}$ depends on the details of the experimental configuration. We find that $I_{thresh}$ is smallest when the pump beams have either linear, orthogonal  or circular, co-rotating polarizations, whereas  $I_{thresh}$ is almost a factor of 2 larger for linearly, co-polarized pump beams. On the other hand, we do not observe superradiance at all for circularly-counterrotating pump beam polarizations, as this configuration does not support atomic bunching.  In addition, we find that superradiance occurs only for $\Delta<0$, which emphasizes the importance of cooling in our geometry \cite{saffman08, tesio12}.  In the remainder of this paper, we focus on the case of linearly cross-polarized pump beams tuned below atomic resonance. 

 Figures \ref{fig:fig1}b and c show the dependence of $I_{thresh}$ on $\Delta$ (for fixed $OD$) and $OD$ (for a fixed $\Delta$), respectively, when we fully extinguish the MOT beams.  We observe that decreasing $|\Delta|$  and increasing $OD$ result in smaller values of $I_{thresh}$,  since these changes enhance the light-matter coupling strength \cite{greenberg11, greenberg12}. The measured values of $I_{thresh}$  agree well with the model developed in Refs. \cite{greenberg11} and \cite{greenberg12} to  describe the light-matter interaction in our system, where we take $I_{thresh}$ as the value of $I_p$ for which the gain becomes infinite.   The lowest threshold we measure is $I_{thresh}=1.1\pm0.25$ mW/cm$^2$, which occurs for $\Delta/\Gamma=-3$ and $OD=20\pm1$ and is comparable to the threshold intensity observed for a Bose-Einstein condensate \cite{inouye99}.   This corresponds to a total input power of $P_{thresh}=2 I_{thresh} \pi (d/2)^2=430\pm80$ $\mu$W in our experiment, where we use pump beams with a $1/e$ diameter of $d=5$ mm.  In contrast to other recently-studied systems, we do not require any auxiliary beams to initially pre-condition the sample \cite{schilke11dfb, zibrov99}.

\begin{figure}
   \begin{center}  \includegraphics[width=3.4in]{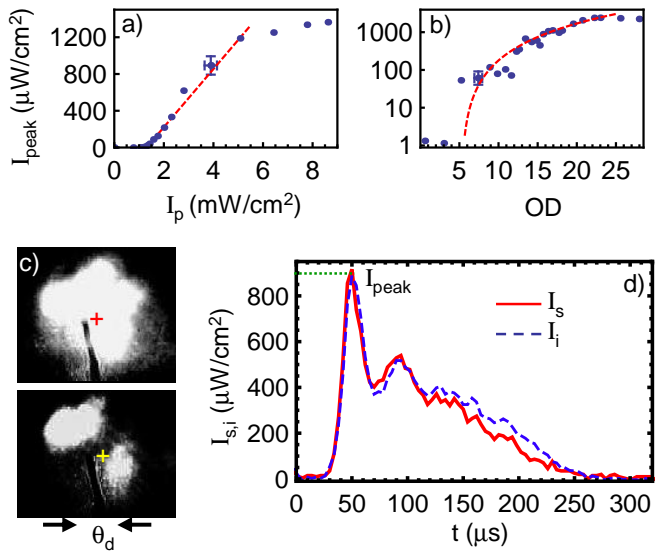}  \end{center} 
   \vspace{-15pt}
 \caption{ Dependence of  $I_{peak}$ on a) $I_p$ for $OD=20\pm1$ and b) $OD$ for $I_p=5\pm0.1$ mW/cm$^2$. The dashed lines in a) and b) are given by the equation $I_{peak}=331(I_p-1.2)$ and $I_{peak}=104(OD-5.2)^2$, respectively, where $I_p$ is in mW/cm$^2$ and $I_{peak}$ is in $\mu$W/cm$^2$. The data in a) and b) correspond to  $\Delta/\Gamma=-5\pm0.5$.   c) Spatial dependence of $I_{s}$ in the far field for two MOT realizations with identical system parameters, where the center of the image  (indicated by a $+$) corresponds to light propagating along the $\hat{z}$-direction. The angular width of each mode is $\theta_D=0.18\pm0.02^\circ$, corresponding to a diffraction-limited beam waist ($1/e$ field radius) of $158\pm20$ $\mu$m at the exit of the trap. The vertical black line is an imaging artifact. d) Temporal dependence of $I_{s,i}$ after we turn the pump beams on at $t=0$.  After a dwell time, the fields grow exponentially, reach a maximum value of $I_{peak}$, and display ringing.  The fields are highly correlated ($r_{s,i}(0)=0.987$) and terminate after $\sim250$ $\mu$s. For the data in c) and d), we use $I_p=4\pm0.1$ mW/cm$^2$, $\Delta=-5\pm0.5$ $\Gamma$ and $OD=30\pm1$.  }
\label{fig:fig2}
\end{figure}

Beyond the instability threshold, the peak intensity of the generated light $I_{peak}$ increases linearly with $I_p$ and quadratically with $OD$, which is a hallmark of collective behavior  (see Figs. \ref{fig:fig2}a and b).  This scaling agrees with previous predictions of superradiance involving atomic bunching \cite{piovella01, slama07}  when one takes into account a finite superradiant threshold \cite{robb04t}. Thus, while we rely on a novel light-matter interaction to achieve superradiance, we find that the above-threshold behavior is largely independent of the details of the underlying gain mechanism.

Mode competition leads to superradiant emission along the direction in which $I_{thresh}$ is smallest.  For our pencil-shaped MOT, we observe the emission of a pair of counterpropagating optical fields  oriented along the vapor's long axis, which we refer to as signal and idler fields with intensities $I_s$ and $I_i$, respectively.  This counterpropagating geometry provides distributed feedback through the mutual coupling of the fields and also balances radiation pressure forces, which is necessary for continuous operation \cite{robb04e}.  Figure \ref{fig:fig2}c shows the measured transverse intensity distribution of the signal beam $I_s(x,y)$ for two different runs with identical system parameters.   The generated light consists of multiple transverse spatial modes, which indicates the existence of atomic self-organization in all three dimensions.  The interplay between mode competition and the random initial  fluctuations that seed the instability causes $I_{s,i}(x,y)$ to vary from run to run \cite{moore99} and represents the spontaneous breaking of continuous translational symmetry in the system.  Our observation of multimode superradiance therefore represents an important step toward realizing novel, controllable condensed matter systems involving phase transitions via emergent structure, and may provide insight into the role of quantum phase transitions in such systems \cite{gopal11, nagy10, strack11}.

The temporal intensity profile of the generated light provides additional insight into the evolution of the instability.   Figure \ref{fig:fig2}d shows $I_{s,i}(t)$, where we turn on the pump beams at $t=0$. As the atoms cool and organize in the presence of the pump beams, the signal and idler intensities  grow exponentially and reach a maximum value $I_{peak}$. The generated fields, which can be as large as $0.2 I_p$, act back on the vapor and strongly affect the ensuing dynamics.  The details of the long-term behavior therefore vary from run to run, but typically consist of additional bursts of light with characteristic durations on the order of $10-100$ $\mu$s due to atomic motion in the underlying optical lattice.  Nevertheless, the signal and idler fields display strong temporal correlations on account of their mutual coupling via the density grating.  We quantify the degree of correlation using the cross-correlation coefficient $r_{s,i}(\tau)=\langle (I_s(t)-\overline{I}_s)(I_i(t+\tau)-\overline{I}_i) \rangle/(\overline{I_s^2}  \overline{I_i^2})^{1/2}$ and find typical values of $r_{s,i}(0)>0.9$. Here, $\langle ... \rangle$ represents a time average and $\overline{A}\equiv \langle A \rangle$.  While we have only measured the classical correlations between the signal and idler beams, we anticipate that quantum correlations also exist and make this system useful for twin beam generation \cite{vallet90}.

\begin{figure}
\begin{center} \includegraphics[width=3.3 in]{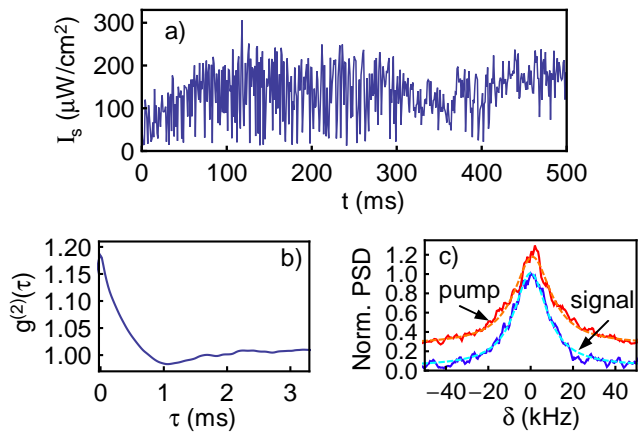} \end{center}
   \vspace{-15pt}
\caption{ a) Typical steady-state temporal waveform of the signal intensity when we leave the MOT beams on at a reduced intensity during the application of the pump beams.  b) The degree of second-order coherence $g^{(2)}_s(\tau)$ for the signal beam time series shown in a).  We find that  $g^{(2)}_{s}(0)=1.18$, where $g^{(2)}(\tau)=1$ for a coherent field. c) Normalized power spectral density (PSD) of the pump-reference (upper, shifted vertically by 0.25) and signal-reference (lower) beatnotes as a function of detuning from the pump frequency ($\delta$).  The solid (dashed) lines correspond to experimental data (Lorentzian fit with a FWHM of 20 kHz).  For the data shown, $I_p=4\pm0.1$ mW/cm$^2$, $\Delta=-6\pm0.5$ $\Gamma$, and  $OD=20\pm1$. }
\label{fig:fig3}
\end{figure}

After a few hundred $\mu$s, loss of atoms from the interaction volume reduces the atomic density and superradiance ends.  We overcome this loss and maintain steady-state superradiance by keeping the MOT beams on at reduced intensities (typically $5-10\%$ of their full values) during the application of the pump beams (see Fig. \ref{fig:fig3}a).  While one might expect random scattering from the MOT beams to inhibit atomic self-organization, we find instead only a slight reduction in the amplitude and degree of correlation ($r_{s,i}(0)\sim0.8$) of the generated light compared to the case of fully-extinguished MOT fields. This situation is analogous to a recently-demonstrated scheme for steady-state superradiance based on the rapid repopulation of a long-lived excited state \cite{meiser10, bohnet12}.  Rather than employing internal states, we use long-lived center-of-mass states and continuously drive atomic bunching more rapidly than the superradiant emission rate. This  scheme enables us to realize superradiance that persists for up to several seconds (limited only by the timing electronics).  

To analyze the steady-state signal, we calculate the degree of second-order coherence  $g^{(2)}_j(\tau)=\langle I_j(t)I_j(t+\tau) \rangle/\langle I_j(t)^2 \rangle$, where  $j=\{s,i\}$.  We find that $g^{(2)}_{s,i}(0)>1$ (\textit{i.e.}, the light is bunched) and decays toward 1 as $\tau\rightarrow \infty$ with a $1/e$  coherence time $\tau_c$ (see Fig. \ref{fig:fig3}b).   For weak pumping (\textit{i.e.}, $I_p\sim I_{thresh}$), we observe that $g^{(2)}_{s,i}(0)\sim1.2$.  In addition,  $\tau_c\sim300$ $\mu$s, which  is over 100 times larger than that observed for a cloud of unconfined atoms at comparable temperatures \cite{greenberg11}.  As we increase $I_p$, $g^{(2)}_{s,i}(0)$ ($\tau_c$) increases (decreases) and approaches a value of 2 (50 $\mu$s). This behavior is consistent with recent predictions on steady-state superradiance \cite{meiser10b}, where the system transitions from the superradiant to thermal regime with stronger pumping.  In addition, because the intensity correlation decays at the collective decay rate while the field correlation decays at the single-particle rate, the superradiant fields may actually be coherent over a time substantially larger   than $\tau_c$ (\textit{e.g.}, by up to a factor proportional to the number of collectively radiating atoms).

This long coherence time corresponds to a narrow linewidth of the emitted light.  To study the spectrum of the superradiant light, we record the beatnote of the signal beam with a separate reference field.  We derive this reference field  from the pump beam laser and  use a pair of identical acousto-optic modulators (AOMs) driven by independent frequency sources to set the frequency different between the pump and reference fields to $\sim1$ MHz.    The observed beatnote spectrum  between the superradiant and reference fields is nearly identical to that of the pump and reference fields (see Fig. \ref{fig:fig3}c): both spectra are centered at the same frequency and are well-described by a Lorentzian profile with a FWHM linewidth of $\sim20$ kHz.  Because this frequency spread is much less than the linewidth of the pump laser ($\sim200$ kHz), we conclude that the superradiant fields have a definite phase relationship with respect to the pump field.    Unfortunately, the relative frequency jitter between the AOM drivers (typically $\sim20$ kHz FWHM) limits our measurement resolution so that we cannot determine the linewidth of the superradiant fields.   Nevertheless, the relative spectral width is nearly a factor of two smaller than the Doppler-broadened gain spectrum observed below the superradiant transition \cite{greenberg11}, which indicates that the coldest atoms participate preferentially in the collective scattering \cite{uys08}.

The unique properties of our system suggest a variety of future applications.  For pump beams focused to an optical cross section (\textit{i.e.}, $\lambda^2/2 \pi$), we predict that we require only $\sim60$ photons to drive the instability.  This low threshold, combined with the system's long coherence time, indicates its potential as a quantum memory \cite{simon07}  or quantum logic element \cite{kimble08}.  In addition, our system's capacity to support multiple spatial modes makes it an excellent candidate for studying continuous-variable quantum information protocols based on spatial multimode entanglement \cite{lassen09, boyer08},  low-light-level all-optical switching \cite{dawes10, greenberg11}, and multidimensional optical soliton formation \cite{fibich07}.  Finally, the self-phase-matching nature of the collective instability may provide a simple path toward the efficient generation of tunable short-wavelength light \cite{robb05}.

We gratefully acknowledge the financial support of the NSF through Grant \#PHY-0855399.


\bibliography{chi5_opex_bibtex}

\begin{thebibliography}{35}
\bibitem [{sfn()}]{sfnote}%
  \BibitemOpen
  \href@noop {} {}\bibinfo {note} {We use the term superradiance in order to be
  consistent with the terminology commonly used in the literature, but the case
  of collective emission from an initially-incoherently prepared sample is more
  accurately called superfluorescence}\BibitemShut {NoStop}%
\bibitem [{wmn()}]{wmnote}%
  \BibitemOpen
  \href@noop {} {}\bibinfo {note} {One can alternatively discuss this gain
  mechanism in terms of multi-wave mixing between the optical and matter
  fields. In this case, superradiance is identical to parametric
  self-oscillation, where the amplification of initial density and optical
  field fluctuations gives rise to macroscopic density gratings and emitted
  optical fields}\BibitemShut {NoStop}%
\end{thebibliography}

\end{document}